# COSMOGONY OF GENERIC STRUCTURES

Thomas Buchert

*Max–Planck–Institut für Astrophysik, D–85740 Garching, Munich, Germany*

## ABSTRACT

The problem of formation of generic structures in the Universe is addressed, whereby first the kinematics of inertial continua for coherent initial data is considered. The generalization to self–gravitating continua is outlined focused on the classification problem of singularities and metamorphoses arising in the density field. Self–gravity gives rise to an internal hierarchy of structures, and, dropping the assumption of coherence, also to an external hierarchy of structures dependent on the initial power spectrum of fluctuations.

## I. Introduction

One of the ambitious goals of modern cosmology is to understand the morphology of large–scale structure as it manifest itself in the distribution of galaxies, but hiding its origin and key–elements of understanding in the distribution of dark matter. A detailed analysis of evolution models for large–scale structure is necessary to uncover the full morphology of the, according to present knowledge, dominating dark matter constituent in the Universe.

The formation and evolution of *generic* structures is another fascinating variation of this theme, supported today by a well–developed mathematical literature on *Lagrange–singularities and metamorphoses* (see: Arnol'd *et al.* 1985, 1988), and sought for in a variety of areas such as for example morphological studies in biology, only to mention one of the fruitful applications of *catastrophe theory* (see: Thom 1975).

Understanding the global morphology of structures with the help of generic singularities that arise locally in the dark matter density field is also the focus of this talk: our aim is to identify the *building blocks* of large–scale structure. The problem here is that the possibility of classifying generic structure elements into a finite set of *morphological germs* is thus far restricted to continua which move under inertia; structure formation in the Universe is, however, a result of self–gravity. We shall approach this problem by first developing the principal picture of the much simpler problem of structure formation in an inertial continuum, then generalizing the framework to self–gravitating continua along the same track of argumentation. Indeed, we shall see that self–gravitating structures can be understood to a large extent on the basis of structures that arise in inertial continua, the gravitational problem, however, reveals large qualitative differences both in the fundamental equations and in the possible richness of morphological manifestations. While the motion of inertial continua is well–understood, the research area of self–gravitating motions is still a sparsely charted terrain with many challenging problems lying ahead.

## II. Structure formation in inertial continua

### II.1. Lagrangian kinematics of inertial continua

In the Lagrangian picture of continuum mechanics we consider the deformation field:

$$\mathbf{x} = \mathbf{f}(\mathbf{X}, t) \quad , \tag{1}$$

where $\mathbf{X}$ are Lagrangian coordinates labelling fluid elements, $\mathbf{x}$ are positions of these elements in Eulerian space at the time $t$, and $\mathbf{f}$ is the trajectory of fluid elements for constant $\mathbf{X}$ (vectors are written bold in this paper). The velocity field of the fluid is the tangential field to this family of curves, $\mathbf{v} := \dot{\mathbf{f}}$; the dot denotes the total (Lagrangian) time–derivative along the curves.

Inertial motion means that the acceleration of fluid elements vanishes along the flow lines, $\mathbf{g} := \ddot{\mathbf{f}} = \mathbf{0}$. Hence, the Lagrangian evolution equations can be written as:

$$\ddot{\mathbf{f}} = \mathbf{0} \quad , \tag{2}$$

which we immediately integrate to obtain the general solution to this problem:

$$\mathbf{f} = \mathbf{X} + \mathbf{V}(\mathbf{X})\,(t - t_0) \quad . \tag{3}$$

The integration constants have been chosen to have the initial positions coinciding with the Lagrangian coordinates, $\mathbf{f}(\mathbf{X}, t_0) =: \mathbf{X}$; the initial velocity field is the second integration constant, $\mathbf{v}(\mathbf{X}, t_0) =: \mathbf{V}(\mathbf{X})$.

The velocity of each fluid element is constant along the flow lines, but can differ for different elements, i.e., the velocity field is in general inhomogeneous. We shall assume throughout most of the following considerations that the initial velocity field is *coherent* on some specified spatial scale on which the structure formation process is studied. This assumption we shall drop at the end of this talk.

### II.2. Generic structure formation in inertial continua

Consider, e.g., a one–dimensional sinusoidal velocity distribution. Continuum elements with higher velocity will catch up those elements with lower velocity and will overtake them. At the instant of overtaking the velocity field in Eulerian space is vertical, after that it will become multi–valued, i.e. three continuum elements arrive at the same Eulerian position. This situation is called *three–stream flow*.

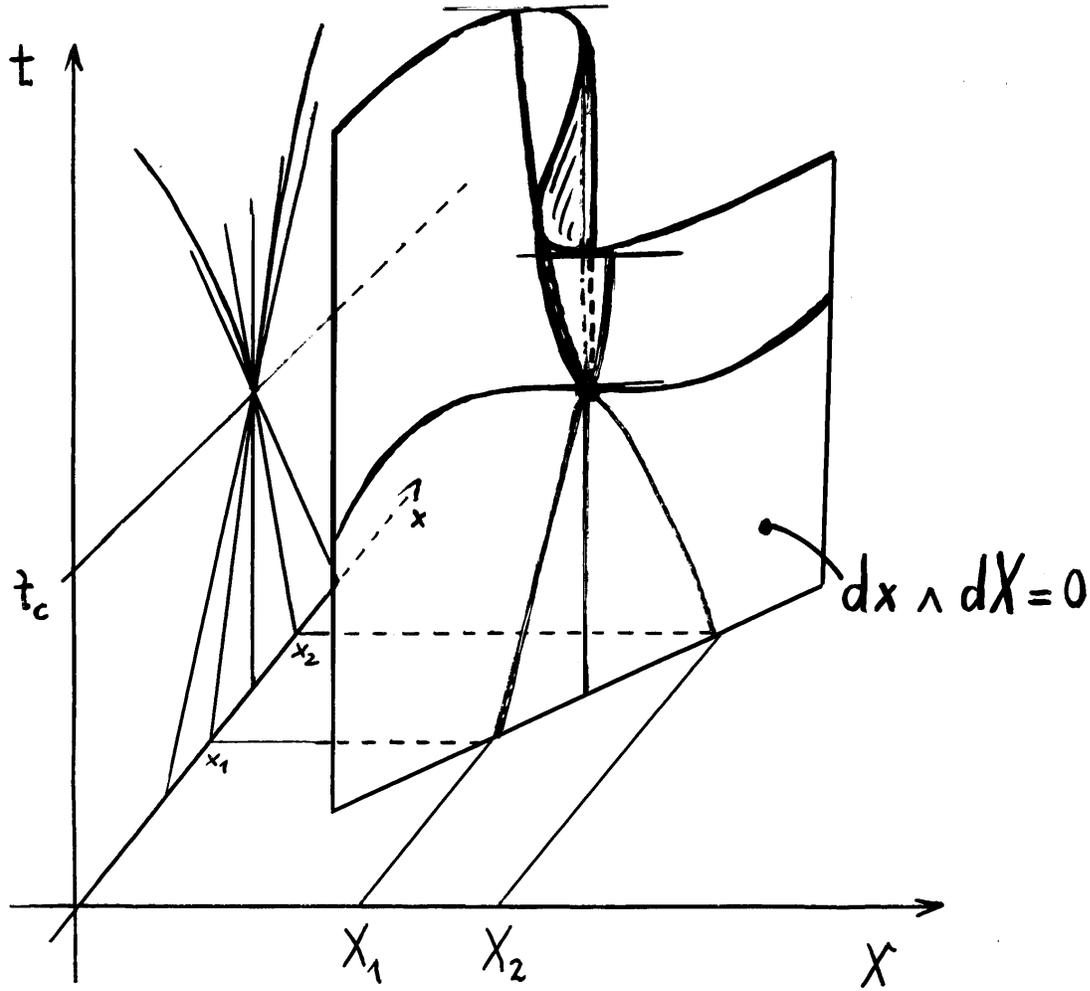

Figure 1: The one–dimensional motion is illustrated in the three–dimensional 'phase space' $(x, X, t)$: trajectories of continuum elements (starting at $x = X$) trace the 'Lagrangian manifold' $\mathcal{L}$ (defined by $dx \wedge dX = 0$). In the case of inertial motion, there is a single folding of $\mathcal{L}$ after shell–crossing; the projection of the critical points of the mapping $X \mapsto x$ form the caustic in Eulerian space enclosing a three–stream flow.

If the initial density field is $\varrho_0(\mathbf{X})$, then the density field in Lagrangian space at the time $t$ changes inversely to the deformation of the continuum. In three dimensions this deformation is measured by the determinant of the deformation tensor $(\partial f_i/\partial X_k)$, and the density in Eulerian space is:

$$\varrho[\mathbf{x}, t] = \sum_{m=1}^{n} \varrho(\mathbf{h}_m[\mathbf{x}, t], t) = \sum_{m=1}^{n} \frac{\varrho_0(\mathbf{h}_m[\mathbf{x}, t])}{|\det[\delta_{ik} + (t - t_0) \frac{\partial V_i}{\partial X_k}(\mathbf{h}_m[\mathbf{x}, t], t)]|} \quad , \qquad (4)$$

where $\mathbf{X}_m = \mathbf{h}_m[\mathbf{x}, t]$ denote the $m$ solutions of the inverse of the transformation (3); $n$ is the number of streams. At the time of verticality of the velocity field, a singularity in the mapping (3) from Lagrangian to Eulerian space develops; images of these singularities lie on surfaces with infinite density in Eulerian space (the *caustic*). Caustics which enclose three–stream flows are called *pancakes* following Zel'dovich (1970) and Arnol'd *et al.* (1982).

The total density in the pancake is the sum of the moduli of the individual densities of the streams according to (4). This is correct as long as the streams do not interact. The one–dimensional picture is illustrated in the space $(x, X, t)$ in Fig.1.

In two–dimensional continua the caustics associated with pancakes have the shape of sickles with cusped edges (compare Arnol'd *et al.* 1982). In order to understand the further evolution of pancakes, we consider the velocity fronts which are at each point orthogonal to the trajectory field. In Fig.2 we illustrate what happens for an initially coherent and anisotropic velocity front. For further discussion of density field metamorphoses the reader may study the articles by Zel'dovich (1970, 1973), and in particular Arnol'd *et al.* (1982) for two–dimensional, and Arnol'd (1982), Buchert *et al.* (1994) for three–dimensional singularities and metamorphoses.

The finite classification of structure elements associated with singularities of the mapping (3) is possible for space dimensions $\leq 5$ (Arnol'd 1982, Arnol'd *et al.* 1985, 1988; see also the contribution of Shenglin Cao (1993) in the Proceedings of our last workshop). However, the mapping (3) defines a one–parameter ($t$) family of Lagrangian mappings only, if the motion is potential, i.e., the mapping (3) from Lagrangian to Eulerian space is a gradient mapping for each $t$. Arnol'd *et al.* (1982) and Buchert *et al.* (1994) give the local algebraic forms of singularities as determined by the eigenvalues of the deformation tensor and its derivatives. This way we are able to predict the formation and transformation of structure elements directly from initial conditions.

The kinematics of an inertial continuum can be understood from the local degeneracies of the deformation tensor: Following Zel'dovich & Myshkis (1973) we write the density field (4) in a diagonal frame at a given point $\mathbf{X}$ for the single–stream potential motion, $\mathbf{V} =: -\nabla_{\mathbf{X}} \mathcal{S}$,

$$\varrho(\mathbf{X}, t) = \frac{\varrho_0(\mathbf{X})}{(1 - \tau \alpha(\mathbf{X}))(1 - \tau \beta(\mathbf{X}))(1 - \tau \gamma(\mathbf{X}))} \quad , \quad \tau = (t - t_0) \quad , \qquad (5)$$

where $\alpha \geq \beta \geq \gamma$ denote the eigenvalues of the tensor $(\partial^2 \mathcal{S} / \partial X_i \partial X_k)$.

In the (non–generic) situation, where all the eigenvalues are equal (spherically symmetric collapse), a continuum element degenerates into a single point. In the generic case, however, the largest eigenvalue $\alpha$ starts to dominate the collapse yielding a strongly anisotropic degeneracy of volume elements into surface elements orthogonal to the eigenvector associated with $\alpha$. After that the second eigenvalue $\beta$ degenerates: the sheet–like element shrinks into a line element along the eigenvector associated with $\beta$. Finally, $\gamma$ degenerates, and the line element shrinks to a point–like matter concentration.

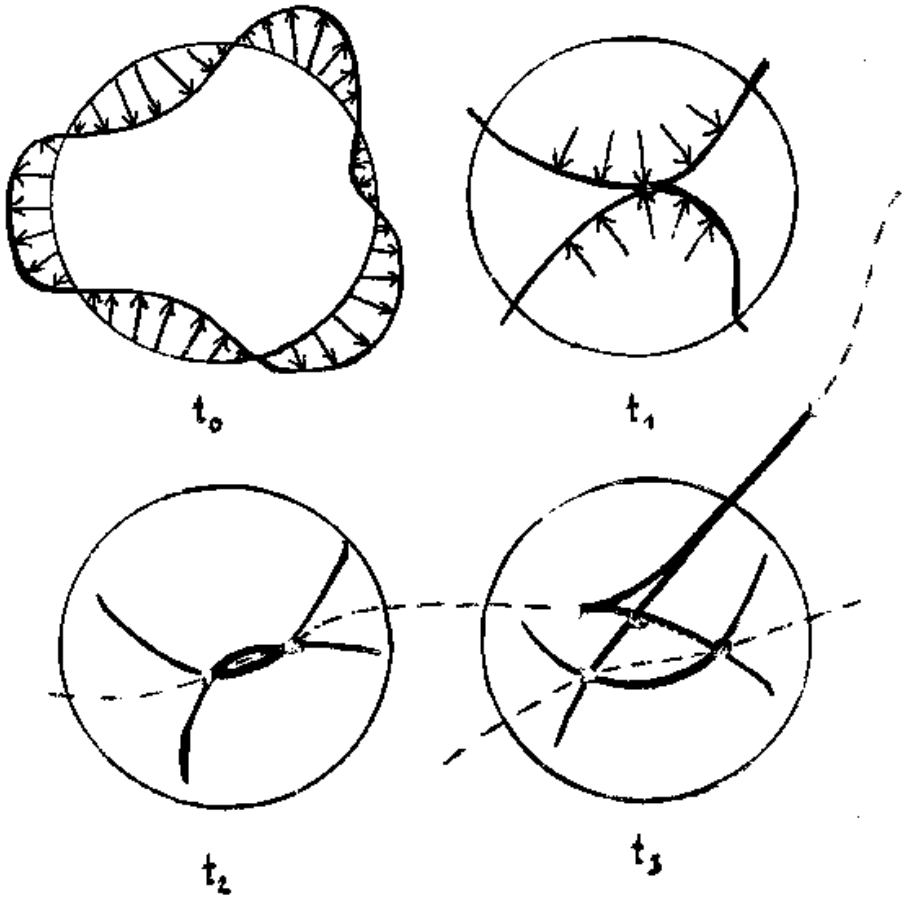

Figure 2: The two–dimensional motion is illustrated in terms of velocity fronts. Initially, the anisotropic deviations of a coherent velocity front from an isotropic (spherically symmetric) front, which would collapse into a single point, is shown. Relative velocity vectors pointing inward move the fronts until they touch each other at a finite instant of time (a 'Legendre singularity' of the front forms corresponding to a 'Lagrange singularity' of the mapping $\mathbf{X} \mapsto \mathbf{x}$). After that time the fronts overlap enclosing a three–stream system of the flow corresponding to a 'pancake' caustic in the density field. The pancake boundary breaks (shown here as the break of the velocity front) with edges connecting other regions of space: this is the origin of a global network structure.

However, this picture is idealized, since after the sheet elements have formed, the mass shells cross freely and form a three–stream system with finite thickness. The principal picture of these subsequent degeneracies is illuminating: it explains locally the existence of flat sheet patterns, filamentary elements and matter concentrations in clumps. If the density field is coherent, this local picture can be extended into the surrounding space, since the eigendirections associated with the eigenvalues only vary slightly with position. Hence, a global picture arises with sheets, filaments and clumps: the *pancake picture* of structure formation (see: Shandarin & Zel'dovich 1989). The density field around such a clump of matter is illustrated in Fig.3.

## III. Structure formation in self–gravitating continua

### III.1. Lagrangian dynamics of self–gravitating continua

The Lagrangian evolution equations (2) are easily generalized to self–gravitating flows:

$$\ddot{\mathbf{f}} = \mathbf{g} \ . \tag{6}$$

Here, the acceleration $\mathbf{g}$ is equivalent to the gravitational field strength according to Einstein's equivalence principle of inertial and gravitational mass. If we restrict ourselves to Newtonian theory, which is adequate for the considered range of scales, then $\mathbf{g}$ is constrained by the field equations: $\nabla_{\mathbf{x}} \times \mathbf{g} = \mathbf{0}$ and $\nabla_{\mathbf{x}} \cdot \mathbf{g} = \Lambda - 4\pi G \varrho$, where the density is again given by the deformation tensor similar to (4); for single–stream systems we have: $\varrho = \varrho_0 J^{-1}$; $J := \det[(\partial f_i / \partial X_k)]$. We have to transform these field equations with respect to Lagrangian coordinates:

$$\epsilon_{pq[j} \frac{\partial(g_{i]}, f_p, f_q)}{\partial(X_1, X_2, X_3)} = 0 \ , \ i \neq j \ , \tag{7a,b,c}$$

$$\sum_{a,b,c} \frac{1}{2} \epsilon_{abc} \frac{\partial(g_a, f_b, f_c)}{\partial(X_1, X_2, X_3)} J^{-1} = \Lambda - 4\pi G \varrho \ . \tag{7d}$$

Inserting (6) and the density integral into (7) we obtain a set of four non–linear partial differential equations for the trajectory field $\mathbf{f}$. Obviously, the gravitational problem is far more complex than inertial motion, which is determined by linear ordinary differential equations for $\mathbf{f}$ (eqs.(2)). (Alternative forms of eqs. (7) are to be found in Ehlers & Buchert 1994.)

Currently, we are not able to give an exact solution to the equations (7) for generic initial data; we can pursue a perturbative analysis and solve the equations (7) up to a given order $r$ with the following perturbation ansatz:

$$\mathbf{f}^{[r]} = a(t)\mathbf{X} + \sum_{\ell=1}^{r} \varepsilon^{\ell} \mathbf{p}^{(\ell)}(\mathbf{X}, t) \ . \tag{8}$$

The first term in this expansion represents the homogeneous and isotropic solutions of the system (7) (Friedmann–Lemaître cosmogonies), in the simplest case $a(t) \propto t^{2/3}$.

Figure 3: The three–dimensional motion is illustrated in terms of an iso–density contour (with low density) at an instant of time, when most of the matter has moved from the sheet pattern towards filaments, and from the filaments towards the clump in the center of the picture (a $D_4$–singularity in the language of the 'Lagrange singularity theory'). In the 'pancake picture' this clump would be identified with the Coma cluster of galaxies embedded into a structure with the morphology of the 'Great Wall'.

## III.2. Relation to the inertial continuum

Going to the first order in the perturbation approach (8) we obtain the following field of trajectories (Buchert 1992):

$$\mathbf{f}^{[1]} = a(t)\,\mathbf{X} + b_1(t)\,\mathbf{U}^D(\mathbf{X})t_0 + b_2(t)\,\mathbf{U}^R(\mathbf{X})t_0 + b_3(t)\,\mathbf{W}(\mathbf{X})t_0^2 \quad, \qquad (9)$$

where $\mathbf{U}^D$, $\mathbf{U}^R$ are the irrotational and rotational parts, respectively, of the initial *peculiar–velocity* (obtained by subtracting the initial *Hubble–velocity* of the homogeneous solution from the total velocity), and similarly $\mathbf{W}$ the *peculiar–acceleration*. A physically interesting class of first–order solutions is provided by the restriction to irrotational flows ($\mathbf{U}^R = \mathbf{0}$), and to parallelity of velocity and acceleration ($\mathbf{U}^D = \mathbf{W}t_0$). This class is singled out by the growing mode solution of the Eulerian linear theory of gravitational instability. With this restriction we recover the well–known *Zel'dovich approximation* (Zel'dovich 1970, 1973); Zel'dovich suggested this extrapolation of the Eulerian linear theory into the weakly non–linear regime:

$$\mathbf{f}^{(Z)} = a(t)\,\mathbf{X} + b(t)\,\mathbf{U}^D(\mathbf{X})t_0 \quad. \qquad (10)$$

The striking fact here is that we understand (10) as a perturbative solution of the full gravitational system (7), and immediately recognize a simple relationship to the inertial continuum studied thus far: the transformation to comoving deformations $\mathbf{q} = \mathbf{F}(\mathbf{X},t) = a(t)^{-1}\mathbf{f}(\mathbf{X},t)$ together with the transformation of the time variable $t \mapsto b(t)/a(t)$ plus the replacement of $\mathbf{V}(\mathbf{X})$ by $\mathbf{U}^D(\mathbf{X})$ shows the formal equivalence of (10) and (3). All statements concerning generic structure elements which are valid in the case of inertial motion also hold to first order in the gravitational problem for this set of initial data.

## III.3. Internal hierarchy of structures

The perturbation approach (8) enables us to study higher–order gravitational effects of the structure formation process. At second order we observe a second shell–crossing singularity which arises as a consequence of the enhanced gravitational self–action of the three–stream system compared to the single–stream system: a second generation of pancakes is created inside the first pancakes (compare Buchert & Ehlers (1993) and my last Proceedings contribution, Buchert (1993), as well as Buchert (1994a) and ref. therein). Numerical N–body simulations show that this is only the onset of a hierarchy of shell–crossings which invoke a cascade of nested pancakes (see: Doroshkevich *et al.* 1980). We already noted that there exist three generations of pancakes due to the degeneracy of three eigenvalues of the deformation tensor in the case of inertial motion.

Self–gravity creates a much richer differentiation of structures, which I call *internal hierarchy* of structures. In a one–dimensional submanifold this is illustrated in Fig.4. Whether this 'self–similar', or 'fractal' substructuring (as suggested by Mandelbrot 1976) continues into a chaotic regime is unclear. There are indications that the first singularity already contains this infinity of bifurcations as is suggested by the study of the second–order solution (see the bifurcation diagram in Buchert & Ehlers 1993), and is supported by the behavior of the gravitational field strength near the singularity: a wide class of solutions to (7) admits 'first integrals' which have the structure that $|\mathbf{g}| \propto \varrho$, i.e., the field strength blows up at caustics. This fact should be compared with the finiteness of $\mathbf{g}$ across caustics as found in Lagrangian perturbation solutions as well as in N–body simulations (see: Buchert 1994b).

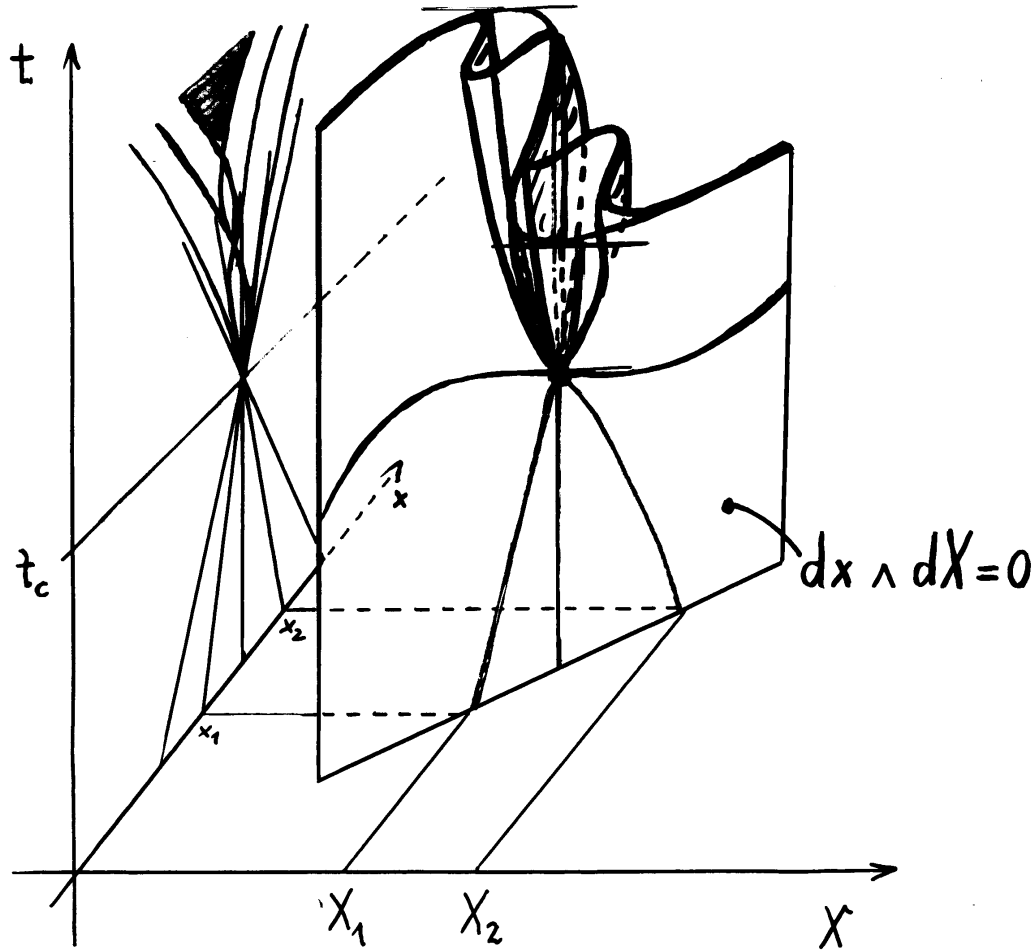

Figure 4: Same as Fig.1, however, in the case of self–gravitating motion, several foldings of $\mathcal{L}$ develop giving rise to several shell–crossings: a $N+2$–stream–system (where $N$ is the number of foldings) is enclosed by a hierarchy of nested caustics in Eulerian space; $N$ is likely to approach $\infty$. Note that here a one–dimensional spatial subsection is presented; the pure one–dimensional case, for which the first–order solution is exact (Buchert 1989), shows a single folding only. (The additional foldings have been added to Fig.1, therefore the single folding is still visible.)

### III.4. External hierarchy of structures

Up to now we studied the structure formation problem for initially coherent fields. Indeed, this picture can be used to successfully describe, at least qualitatively, present day observational data of the galaxy distribution: Identifying the central peak in Fig.3 with, e.g., the Coma cluster of galaxies, then the filamentary connections between clusters and even sheet patterns can be found in the observations (the so–called *Great Wall*, Geller & Huchra 1989). However, since sheet patterns have comparatively low densities, one usually only observes filamentary and clumpy patterns in the galaxy distribution, the sheets being hidden in the unseen dark matter constituent.

There is considerable disagreement in the community as to whether this simplified picture is applicable; in other words, it is doubtful that such a coherence scale is present in the initial data at the recombination epoch. However, a pure baryonic universe as well as a 'Hot–Dark–Matter' universe show such coherence properties as a result of viscous, or free–streaming damping mechanisms, respectively.

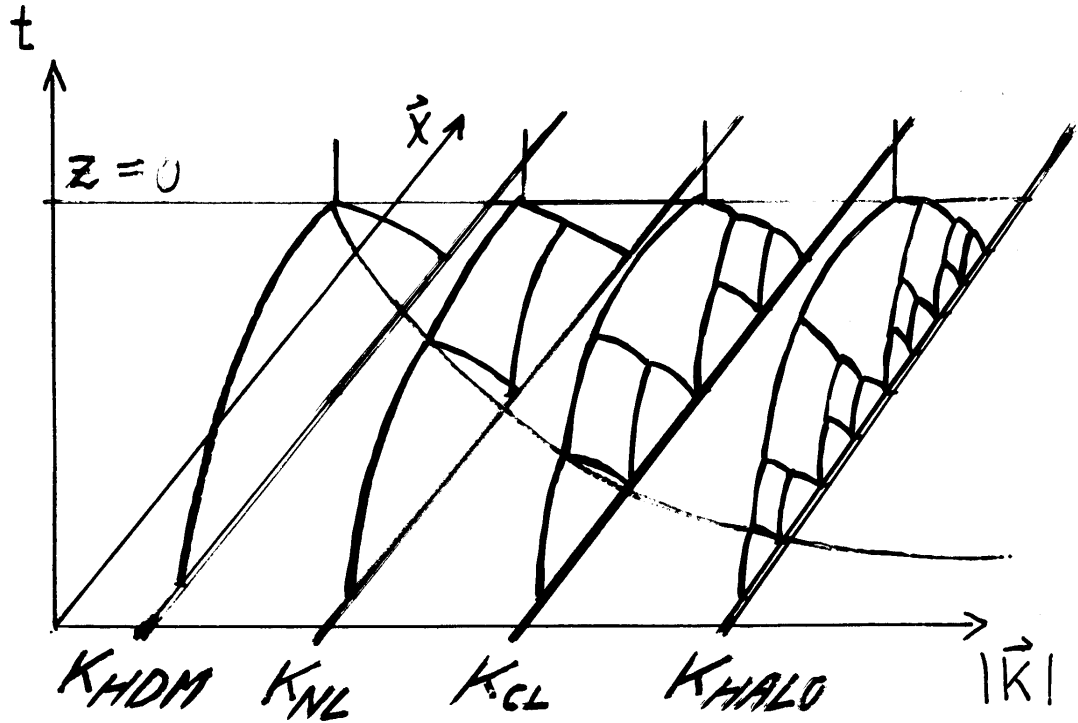

Figure 5: A scetch of the structure formation process in hierarchical models: Different spatial scales are marked; the collapse process is symbolized by crossing trajectories. Shell–crossing occurs earlier on smaller spatial scales. However, the merging of collapsed objects is controlled by the hierarchically superordered scale in such a way that the collapsed objects are "guided" by the collapse of the larger wavelength. Thus, truncation of higher frequencies in the initial power spectrum will not influence the structure formation process significantly on larger scales, as was demonstrated by Melott et al. 1994a,b.

In order to study non–coherent initial data, we choose as description their power spectrum $\mathcal{P}(k)$, i.e., the ensemble average of the squared absolute values of the Fourier coefficients of the density field. For simplicity, we shall discuss cases in which the power spectrum has the form of a powerlaw: $\mathcal{P} = k^n$. The index $n = -3$ here plays a special role; it is that index for which the total power in a given $k$–interval, $\int d^3k\ \mathcal{P}(k)$, is approximately constant (in fact logarithmic) as a function of scale $k$.

In such a model, the self–similar linear growth of fluctuations implies that all scales enter the non–linear regime at about the same time. Consequently, all the structures we described in the present talk form on all scales simultaneously. For less small–scale power, we recover those models which have a sufficiently steep fall–off in the spectrum at short wave–lengths, thus forming structures on a scale with most power (usually the maximum of the spectrum like in a 'Hot–Dark–Matter' cosmogony). In the opposite case, we encounter those models which commonly are called *hierarchical*, i.e., structures form first on smaller scales and then merge into superunits on larger scales (as an example: the 'Cold–Dark–Matter' cosmogony). Especially the latter models result in that scale–dependent phenomenology which I call *external hierarchy* of structures (scetched in Fig.5). Such models, in particular those which are close to $n = -3$ on small scales, are the most likely ones in comparison with observational data. That the picture of generic structure metamorphoses I described in this talk is applicable also in the case of hierarchical cosmogonies has been demonstrated by Melott *et al.* (1994a,b); see also the discussion in Buchert *et al.* (1994).

## IV. Open problems

Summarizing, I have presented a picture which is characterized by structure metamorphoses on all spatial scales with a continuing hierarchical feature of the internal structure differentiation down to smaller and smaller scales.

This picture is limited on large scales by the horizon scale, where general relativistic models for self–gravitating motion have to be considered; on small scales, which are not ruled by the gravitational interaction, hydrodynamical effects of the baryonic matter and chaotic thermal velocities of collisionless matter are relevant: still, self–gravitating structures idealized by images of generic singularities are thought to form the skeleton of large–scale structure down to the galaxy as structural unit. The gravitational interaction of multi–stream systems is an open problem in the description of structure formation behind caustics: the basic equations (7) have to be extended to the Vlasov–Poisson system in order to properly meet this problem.

Another unsolved problem is the classification of generic structure elements in the case of self–gravitating motion: There is no proof specifying that class of motions which develop Lagrangian singularities. For general motions the mathematical background has to be extended.

Since we observe hierarchical features on all spatial scales, it is not at all intruding that the universal structure admits a *scale of homogeneity*, i.e., that the power spectrum bends over to a falling slope on the largest scales. It is likely that this hierarchy, which we observe in a large window of spatial scales, continues also to scales which are beyond the access of current observations. Indications from the COBE measurement of the

large–scale end of the power spectrum that this is not the case I consider as preliminary: cosmic variance arguments express the statistically incomplete status of this observation (see: Scaramella & Vittorio 1993). Both *internal* and *external* hierarchies may be an indication that we have to develop a fully hierarchical model of the universe abandonning the prejudice that the universe should follow the Friedmann–Lemaître cosmogonies on average and that it should admit scale–independent average characteristics (like the Hubble constant) calculated from these simple–minded models.

**Acknowledgements:**

*I would like to thank Matthias Bartelmann for providing Fig.3 as well as Thomas Boller for scanning the figures. I am greatful to Profs. Xiaoyang Xia and Zugan Deng for their efforts to make this workshop most enjoyable.*